\documentclass[conference]{IEEEtran}
\usepackage{amssymb,stmaryrd,amsmath,amsfonts,rotating}
\usepackage{color}
\usepackage{bm}
\usepackage{latexsym}
 \newcommand{\qed}{\hfill \ensuremath{\Box}}

\usepackage{mathtools}
\mathtoolsset{showonlyrefs}

\newcommand{\ei}{\ensuremath{\epsilon_{\underline{i}}}}
\newcommand{\ZL}{\ensuremath{\mathbb{Z}_L}}
\newcommand{\ZZ}{\ensuremath{\mathcal{Z}}}

\newtheorem{theorem}{Theorem}
\newtheorem{definition}{Definition}

\newtheorem{lemma}{Lemma}
\newtheorem{discussion}{Discussion}
\newtheorem{proposition}{Proposition}

\newcommand{\dl}{d_{{l}}}

\newcommand{\dr}{d_{{r}}}

\allowdisplaybreaks
\usepackage{graphicx}	
\begin{document} 
\title{Multi-Dimensional Spatially-Coupled Codes}
\author{
  \IEEEauthorblockN{Ryunosuke Ohashi and Kenta Kasai}
  \IEEEauthorblockA{
Dept. of Commun. \& Integraded Systems,\\
Tokyo Institute of Technology, \\
152-8550 Tokyo, Japan.\\
Email: 
{\tt \{ohashi8,kenta\}@comm.ss.titech.ac.jp},
} 
  \and
  \IEEEauthorblockN{Keigo Takeuchi}
  \IEEEauthorblockA{
Dept. of Commun. Engineering \& Inf.\\
University of Electro-Communications\\
Tokyo 182-8585, Japan.\\
Email:{\tt \{ktakeuchi\}@uec.ac.jp}
}
}
\maketitle
\begin{abstract}
Spatially-coupled (SC) codes are constructed by coupling many regular low-density parity-check codes in a chain. 
The  decoding chain of SC codes stops when facing burst erasures. 
This problem can not be overcome by increasing coupling number. 
In this paper, we introduce multi-dimensional (MD) SC codes. 
Numerical results show that 2D-SC codes are more robust to the burst erasures than 1D-SC codes. 
Furthermore, we consider designing MD-SC codes with smaller rateloss. 
\end{abstract}
\section{Introduction}
Spatially-coupled (SC) low-density parity-check (LDPC) codes attract
 much attention due to their capacity-achieving performance and a memory-efficient sliding-window decoding algorithm.
The studies on SC-LDPC codes date back to the invention of convolutional LDPC codes by Felstr{\"o}m and Zigangirov \cite{zigangirov99}. 
Lentmaier {\it et al.}~ observed that (4,8)-regular convolutional LDPC codes exhibited the decoding performance surpassing the belief propagation (BP) threshold of (4,8)-regular block LDPC codes \cite{5571910}. 
Further, the BP threshold coincides with the maximum a posterior (MAP) threshold of the underlying block LDPC codes with a lot of accuracy. 
Constructing convolutional LDPC codes from a block LDPC code improves the BP threshold up to the MAP threshold of the underlying codes. 

Kudekar {\it et al.}~ named this phenomenon  ``threshold saturation'' and proved rigorously for the binary-input erasure channel (BEC) \cite{5695130} and the binary-input memoryless output-symmetric (BMS) channels. \cite{2012arXiv1201.2999K}. 
In the limit of large $\dl,\dr,L$ and $w$, the SC-LDPC code ensemble $(\dl,\dr,L,w)$ \cite{5695130} 
was shown to {\em universally} achieve the Shannon limit of the binary-input memoryless symmetric-output (BMS) channels under BP decoding. 

In this paper, we deal with a serious problem of SC-LDPC codes. 
SC-LDPC codes are constructed by coupling $L$ regular LDPC codes in a chain. 
Belief propagation (BP) is employed to decode the chain of codes starting from the end points of the chain. 
The BP decoding of SC codes stops when facing burst erasures. 
In other words, the decoding error probability remains positive  from the section at which the burst erasures are received. 
This problem can not be solved by increasing coupling number. 
In this paper, we introduce multi-dimensional (MD) SC codes to overcome this problem. 
Numerical results show that 2D-SC codes are more robust to the burst erasures than 1D-SC codes. 
Furthermore, we consider designing MD-SC codes with small rateloss as $O(1/L^D)$, where $D$  is the dimension.
\section{Multi-Dimensional Coupled Codes}
\subsection{Definition: $(\dl,\dr,L,\omega,\ZZ)$ codes}
\begin{definition}
 Define $\ZL:=\mathbb{Z}/L\mathbb{Z}=\{0,1,\dotsc,L-1\}$. For $a,b\in \ZL$.
 Consider $L^D$ sections on $D$-dimensional discrete torus $\ZL^D$. 
 For bit node degree $\dl\ge 3$ and check node degree $\dr> \dl$, 
a coupling number $L>w$,  connecting rate $0\le \omega_{\underline{j}}\le 1\ (\underline{j}\in \ZL^D)$, and a shortened domain $\ZZ\subset \ZL^D$, 
we define ML-SC $(\dl,\dr,L,\omega, \ZZ)$ codes as follows. 
Throughout this paper, we fix ${\omega}_{\underline{j}}$ as
\begin{align}
  &{\omega}_{\underline{j}}=
  \begin{cases}
   1/w^D&\underline{j}\in [0,w-1]^D\\
   0&\mbox{otherwise}, 
  \end{cases}
\end{align}
where we denoted $[a,b]:=\{a,a+1,\dotsc,b-1,b\}$. 

 Each section $\underline{i}\in[0,L-1]^D$ has $M$ bit nodes of degree $\dl$ and
 $\frac{\dl}{\dr}M$ check nodes of degree $\dr$. 
 Connect edges between bit nodes and check nodes uniformly at random so that bit nodes 
in section $\underline{i}$ are 
connected to check nodes in section $\underline{i}+\underline{j}\ (\underline{j}\in \ZL^D)$ with 
$\omega_{\underline{j}} M$ edges, respectively. 
Shorten the bit nodes in section $\underline{i}\in \ZZ\subset\ZL^D$. 
Namely, the shortened bit nodes are set to 0 and are not transmitted through the channel. 
\label{234326_24Jan13}
\end{definition}
\begin{discussion}
In \cite{5695130}, spatially-coupled codes of coupling number $L$ were defined over section $[-L,+L]$ and the bit nodes outside $[-L,+L]$ were shortened. 
Some might think it is more natural to define MD-SC codes over $[-L,+L]^D$ than over $[0,L-1]^D$ and shorten the bit nodes outside $[-L,+L]^D$. 
If we defined so, it would be difficult to distinguish the effect of MD extension from the boundary effect from each dimension as 1D-SC codes. 
This is why we employ the codes in Definition \ref{234326_24Jan13}. 
\end{discussion}
\begin{lemma}
The coding rate $R(\dl,\dr,L,\omega,\ZZ)$ of $(\dl,\dr,L,\omega,\ZZ)$ codes is given by 
 \begin{align}
& 1-\frac{\dl}{\dr}\frac{1}{L^D-\#\ZZ}\sum_{\underline{i}\in\ZL^D}\Bigl(1-\bigl(\sum_{\underline{j}:\underline{i}+\underline{j}\in \ZZ}\omega_{\underline{j}}\bigr)^{\dr}\Bigr).\label{144620_27Jan13}
 \end{align}
\end{lemma}
{\itshape Proof}:
We will count the number of transmitted bit nodes and valid check nodes. 
Let $V$ and $C$ denote these numbers, respectively.
Since check nodes adjacent only to shortened bit nodes are not giving any constraint on the code, 
it is sufficient to count  the check nodes adjacent to unshortened bit nodes. 
Since the degree of check nodes are $\dr$, a check node in section $\underline{i}$ has $\dr$ edges connecting to shortened bit nodes with probability
\begin{align}
 \bigl(\sum_{\underline{j}: \underline{i}+\underline{j}\in \ZZ}\omega_{\underline{j}}\bigr)^{\dr}. 
\end{align}
Therefore, the average number of check nodes which are adjacent to at least one unshortened bit nodes is given by 
\begin{align}
C=\frac{\dl}{\dr}M\sum_{\underline{i}\in\ZL^D}\Bigl(1-\bigl(\sum_{\underline{j}: \underline{i}+\underline{j}\in \ZZ}\omega_{\underline{j}}\bigr)^{\dr}\Bigr). 
\end{align}
There are $V=M(L^D-\#\ZZ)$ unshortened bit nodes. We calculate the coding rate  as $1-C/V$, which concludes the lemma. \qed
\section{Density Evolution Analysis}
We consider the transmission takes place over the BEC($\epsilon$) with erasure probability $\epsilon$. 
The BP decoding is employed. 
Let $p_{\underline{i}}^{(\ell)}$ denote the erasure probability of BP messages from bit nodes to check nodes at the $\ell$-th iteration round. 
Let $q_{\underline{i}}^{(\ell)}$ denote the erasure probability of BP messages from check nodes to bit nodes at the $\ell$-th iteration round. 
Since the bit nodes in section in $\ZZ$ are shortened, $p_{\underline{i}}^{(0)}$ are given as
\begin{align}
p_{\underline{i}}^{(0)}=\ei :=
\begin{cases}
 0 &(\underline{i}\in \ZZ),\\
 \epsilon\ &(\underline{i}\notin \ZZ).
\end{cases}
\end{align} 
For $\ell\ge 1$,  $p_{\underline{i}}^{(\ell)}=0$ for shortened section $\underline{i}\in \ZZ$
and 
\begin{align}
 &  p_{\underline{i}}^{(\ell)}=\ei\Bigl(\sum_{\underline{j}} \omega_{\underline{j}}q_{\underline{i}+\underline{j}}^{(\ell)}\Bigr)^{\dl-1}\label{165410_2Jan13},\\
 &  q_{\underline{i}}^{(\ell)}=1-\Bigl(1-\sum_{\underline{j}}\omega_{\underline{j}}p_{\underline{i}-\underline{j}}^{(\ell-1)}\Bigr)^{\dr-1}, 
\end{align}
for $\underline{i}\notin \ZZ$.
The decoding erasure probability $\mathbb{P}_\mathrm{b}^{(\ell)}$ is given by 
\begin{align}
\mathbb{P}_\mathrm{b}^{(\ell)} &=\frac{1}{L^D-\#\ZZ}\sum_{\underline{i}\in\ZL^D}\ei\Bigl(\sum_{\underline{j}}
 \omega_{\underline{j}}q_{\underline{i}+\underline{j}}^{(\ell)}\Bigr)^{\dl}. 
\end{align}
We define the BP threshold $\epsilon^{\ast}$ as
\begin{align}
\epsilon^{\ast}= \sup\{\epsilon>0\mid \lim_{\ell\to\infty}\mathbb{P}_\mathrm{b}^{(\ell)}=0\}. 
\end{align}
Namely,  for the erasure probability below the threshold $\epsilon^{\ast}$, the decoding erasure probability goes to zero. 
\subsection{Shortening hyperplane of width $w$}
Choose a hyperplane of width $w$ in $D$-dimensional space is chosen as shortened domain ${\ZZ}$. 
The following proposition asserts that the density evolution is equivalent to that of  of 1D system. 
\begin{proposition}
Let us define
 \begin{align}
  &\tilde{\ZZ}:=[0, w-1],\\
  &{\ZZ}:=\{\underline{i}=(i_1,\dotsc,i_D)\in \ZL^D\mid i_D\in [0,w-1]\}.
\end{align}
We use $\tilde{\cdot}$ for variables for 1D system throughout this paper for the sake of readability. 
Then we have 
\begin{align}
\tilde{\epsilon}^{\ast}(\dr,\dr,L,\tilde{\omega},\tilde{\ZZ})&=\epsilon^{\ast}(\dr,\dr,L,{\omega},{\ZZ}),\label{233750_20Jan13}\\
 \tilde{R}(\dr,\dr,L,\tilde{\omega},\tilde{\ZZ})&=R(\dr,\dr,L,{\omega},{\ZZ})\label{233502_20Jan13}\\
&=1-\frac{\dl}{\dr}-O(1/L).\label{011145_25Jan13}
 \end{align}
\end{proposition}
{\itshape Proof}:
We give a proof for $D=2$. The proof for $D>2$ follows similarly. 
It is sufficient to show $\tilde{p}_{i_1}^{(\ell)}={p}_{\underline{i}}^{(\ell)}$ for any $\ell\ge0$.
It is obvious that 
\begin{align}
 \tilde{p}_{i_1}^{(0)}={p}_{(i_1,i_2)}^{(0)} =
\begin{cases}
0 & i_1\in [0,w-1]\\
\epsilon &\mbox{otherwise}, 
\end{cases}
\end{align}
for $i_1,i_2\in \ZL$. 
Assume $ \tilde{p}_{i_1}^{(\ell)}={p}_{(i_1,i_2)}^{(\ell)}$ for $\ell$. 
From Eq.~\eqref{165410_2Jan13} and the definition of $\tilde{\omega}$ and ${\omega}$, it follows that 
\begin{align}
  {q}_{(i_1,i_2)}^{(\ell+1)}&=1-\Bigl(1-\frac{1}{w^2}\sum_{j_1=0}^{w-1}\sum_{j_2=0}^{w-1} {p}_{(i_1-j_1,i_2-j_2)}^{(\ell)}\Bigr)^{\dl-1}\\
 & =1-\Bigl(1-\frac{1}{w}\sum_{j_1=0}^{w-1}\tilde{p}_{i_1-j_1}^{(\ell)}\Bigr)^{\dl-1}= \tilde{q}_{i_1}^{(\ell+1)},\\
   {p}_{(i_1,i_2)}^{(\ell+1)}&=\ei\Bigl(\frac{1}{w^2}\sum_{j_1=0}^{w-1}\sum_{j_2=0}^{w-1} {q}_{(i_1+j_1,i_2+j_2)}^{(\ell+1)}\Bigr)^{\dl-1}\\
 &  =\epsilon\Bigl(\frac{1}{w}\sum_{j_1=0}^{w-1}\tilde{q}_{(i_1+j_1)}^{(\ell+1)}\Bigr)^{\dl-1}=\tilde{p}_{i_1}^{(\ell+1)}. 
\end{align}
Thus we have $\tilde{p}_{i_1}^{(\ell)}={p}_{(i_1,i_2)}^{(\ell)} $ for any $\ell\ge 0$, which concludes \eqref{233750_20Jan13}. 
We derive \eqref{233502_20Jan13} as follows. 
\begin{align}
&R(\dr,\dr,L,{\omega},{\ZZ})\\
&=  1-\frac{{\dl}/{\dr}}{L^2-wL}\sum_{(i_1,i_2)\in\ZL^2}\Bigl(1-\bigl(\sum_{(j_1,j_2):(i_1+j_1,i_2+j_2)\in \ZZ}\hspace{-0.8cm}\omega_{(j_1,j_2)}\bigr)^{\dr}\Bigr)\\
&=  1-\frac{{\dl}/{\dr}}{L^2-wL}L\sum_{(i_1,0)\in\ZL^2}\Bigl(1-\bigl(\sum_{(j_1,j_2):(i_1+j_1,j_2)\in \ZZ}\hspace{-0.8cm}\omega_{(j_1,j_2)}\bigr)^{\dr}\Bigr)\\
 &=  1-\frac{{\dl}/{\dr}}{L-w}\sum_{(i_1,0)\in\ZL^2}\Bigl(1-\bigl(\sum_{j_2=0}^{w-1}\sum_{j_1:0\le i_1+j_1\le w-1}\hspace{-0.8cm}1/w^2\bigr)^{\dr}\Bigr)\\
&=R(\dr,\dr,L,\tilde{\omega},\tilde{\ZZ}). 
\end{align}
Equation \eqref{011145_25Jan13} follows from 
\begin{align}
 R(\dr,\dr,L,\tilde{\omega},\tilde{\ZZ})=\Bigl(1-\frac{\dl}{\dr}\Bigr)-\frac{\dl}{\dr}\frac{1-w-2\sum_{i=0}^w(\frac{i}{w})^{\dr}}{L-w}
\end{align}
of which proof is appeared in \cite{5695130} for coupled codes defined on $[-L,L]$. 
\qed

In the next section, we will see these MD-SC codes with shortening domain as a hyperplain behave differently from the 1D-SC codes. 
\section{Robustness for Burst Erasure}

In this section, we consider burst erasures and demonstrate robustness of 2D coupled codes. 
Spatially coupled codes are constructed by coupling $L$ regular LDPC codes of length $M$. 
Assume that we are transmitting bits coded by 1D-SC codes and a burst section erasure of length $M$ occur at section $i$. 
We call such burst erasures for a section a {\itshape burst section erasure}. 
Decoding proceeds from the section in $\ZZ$. Such burst section erasure is described as $\epsilon_{\underline{i}}=\epsilon_i=1$. 

\begin{figure}[t]
\begin{picture}(200,170)(0,0)
\put(00,5){   \includegraphics[width=0.48\textwidth]{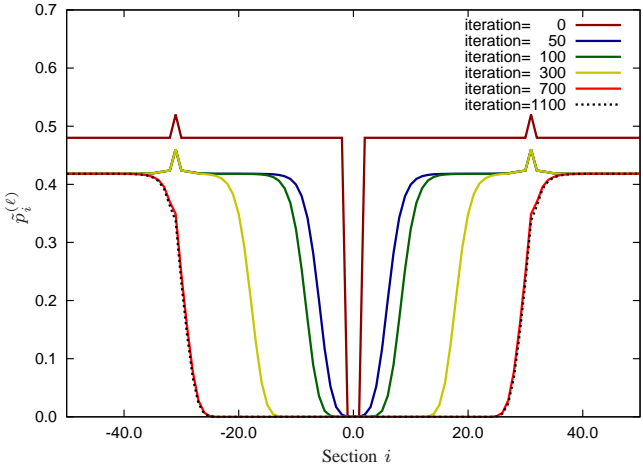}}
\put(0,90){\rotatebox{90}{\scriptsize{$\tilde{p}_i^{(\ell)}$}}}
\put(120,0){\scriptsize{Section $i$}}
\end{picture}
\caption{Transition of message error probability $\tilde{p}_i^{(\ell)}$ at each section $i$ and at iteration $\ell$ of 1D-SC $(\dl=3,\dr=6,L=101,\tilde{\omega},\tilde{\ZZ}=\{0,\pm 1\})$ codes with $w=4$. 
 The channel is  BEC($\epsilon=0.48$) with  2 burst section erasures injected. 
Decoding stops around at section $i=\pm 31$ where burst erasures are injected as $\tilde{p}_i^{(0)}=0.52. $}
\label{212717_27Jan13}
\end{figure}
Can 1D-SC codes correct such burst erasures?
Figure \ref{212717_27Jan13} shows  the transition of message error probability of 1D-SC $(\dl=3,\dr=6,L=101,\tilde{\omega},\tilde{\ZZ}=\{0,\pm 1\})$ codes with $w=4$. 
 The channel is  BEC($\epsilon=0.48$) with  2 burst section erasures injected. 
Decoding stops around at section $i=\pm 31$ where burst erasures are injected as $\tilde{p}_i^{(0)}=0.52$. The 1D-SC codes can not recover such burst erasures. 

\begin{figure*}
\begin{picture}(400,270)(-10,0)
\put(  0,180){\includegraphics[width=0.245\textwidth]{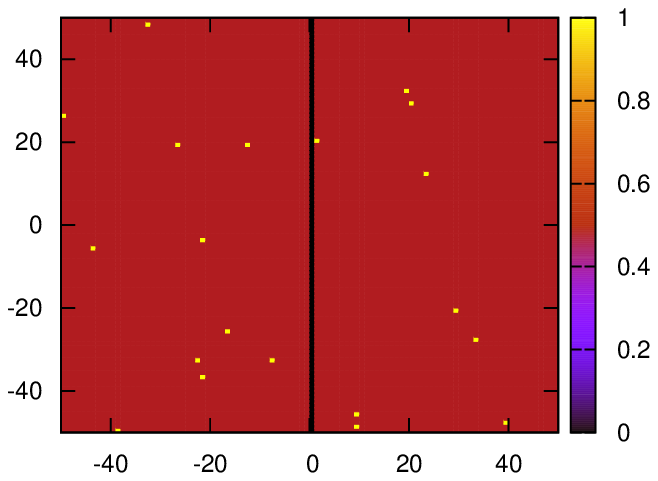}}
\put(120,180){\includegraphics[width=0.245\textwidth]{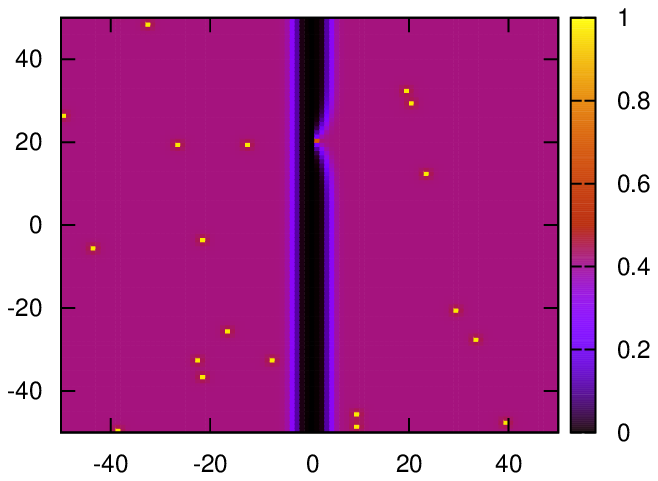}}
\put(240,180){\includegraphics[width=0.245\textwidth]{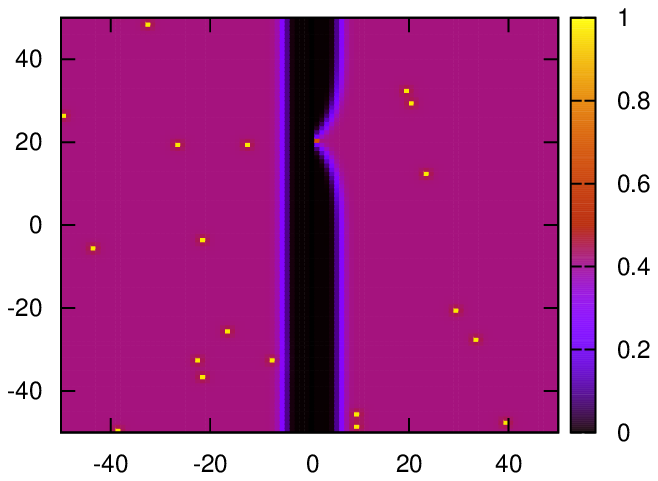}}
\put(360,180){\includegraphics[width=0.245\textwidth]{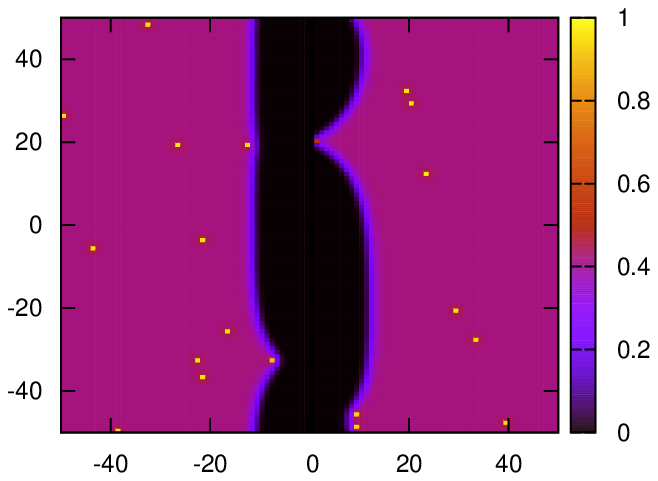}}
\put(  0, 90){\includegraphics[width=0.245\textwidth]{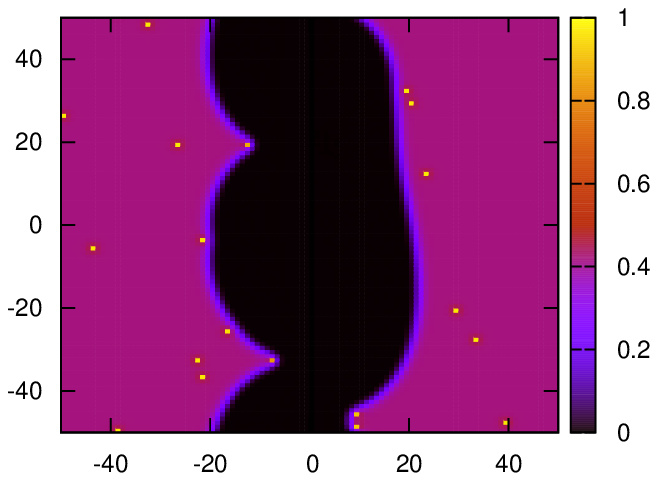}}
\put(120, 90){\includegraphics[width=0.245\textwidth]{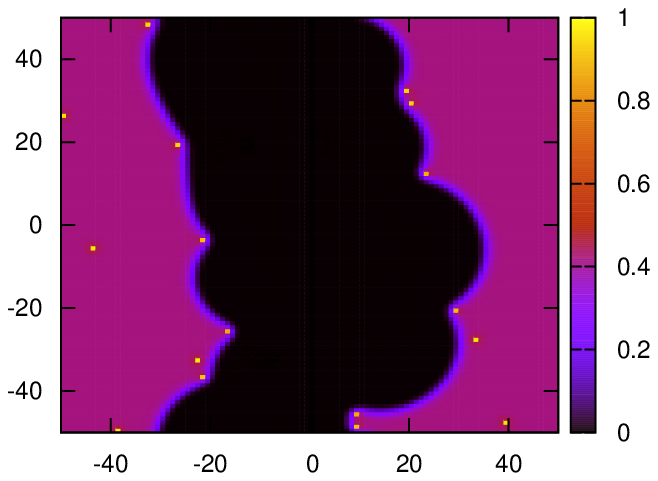}}
\put(240, 90){\includegraphics[width=0.245\textwidth]{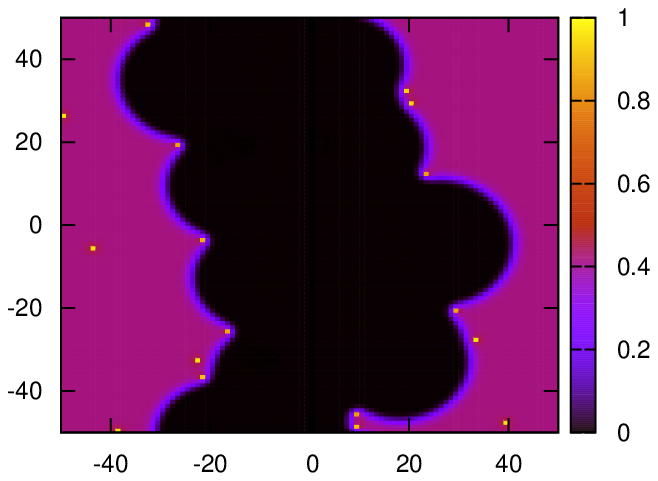}}
\put(360, 90){\includegraphics[width=0.245\textwidth]{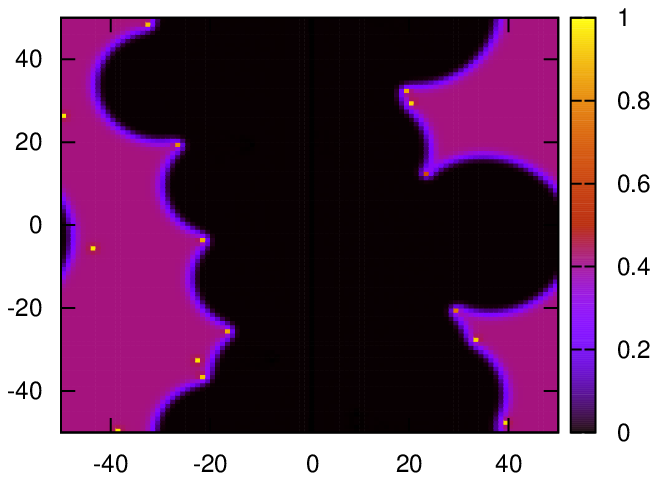}}
\put(  0,  0){\includegraphics[width=0.245\textwidth]{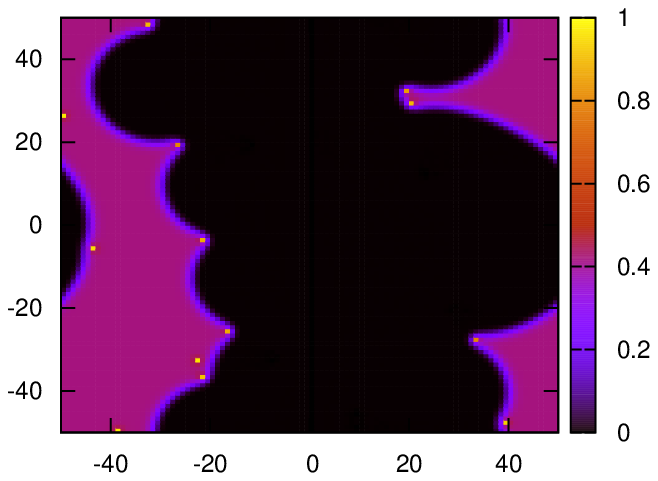}}
\put(120,  0){\includegraphics[width=0.245\textwidth]{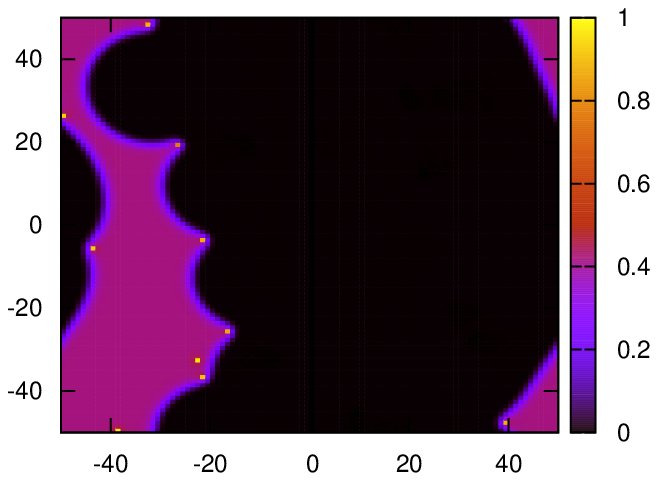}}
\put(240,  0){\includegraphics[width=0.245\textwidth]{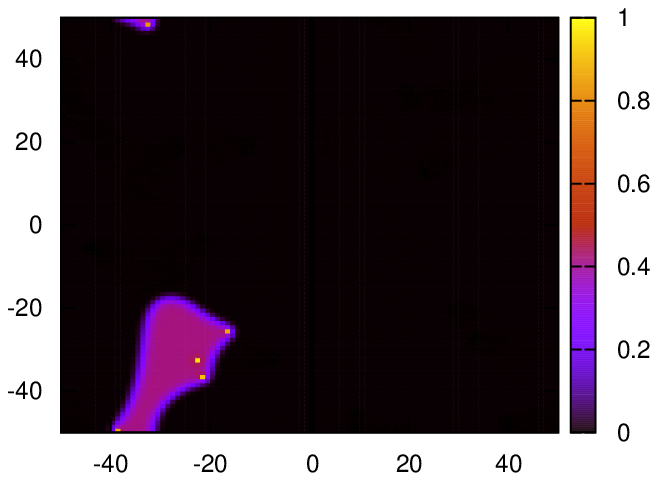}}
\put(360,  0){\includegraphics[width=0.245\textwidth]{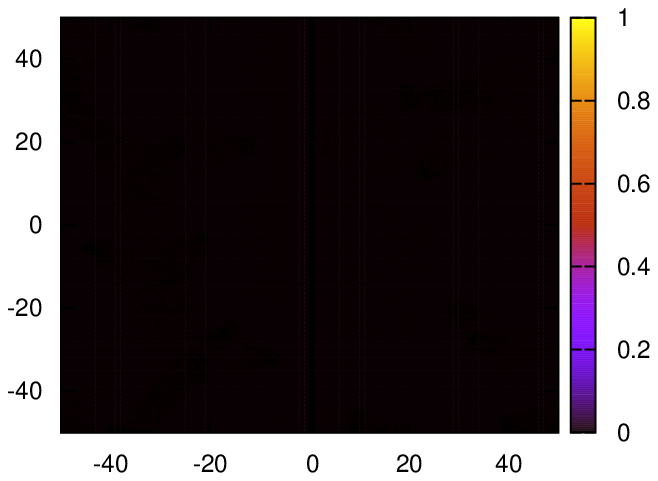}}
\end{picture} 
\caption{Transition of decoding error rate of 2D-SC $(\dl=3,\dr=6,L=101,\omega,\ZZ)$ codes with $w=2$ and  $\ZZ$ as line segment of width 1.  
 The channel is  BEC($\epsilon=0.48$) with  20 burst section erasures injected. }
\label{213143_27Jan13}
\end{figure*}

Figure \ref{213143_27Jan13}, shows 
the transition of decoding error rate of 2D-SC $(\dl=3,\dr=6,L=101,\omega,\ZZ)$ codes with $w^2=4$ and  $\ZZ$ as square segment of size 15. 
 The channel is  BEC($\epsilon=0.48$) with  20 burst section erasures injected. Each burst section erasures are described as $p^{(0)}_{\underline{i}}=1.0$. 
The 2D-SC codes are capable of recovering such burst section erasures. 

\begin{figure}[t]
\begin{picture}(260,175)(0,0)
\put(10,5){\includegraphics[width=0.48\textwidth]{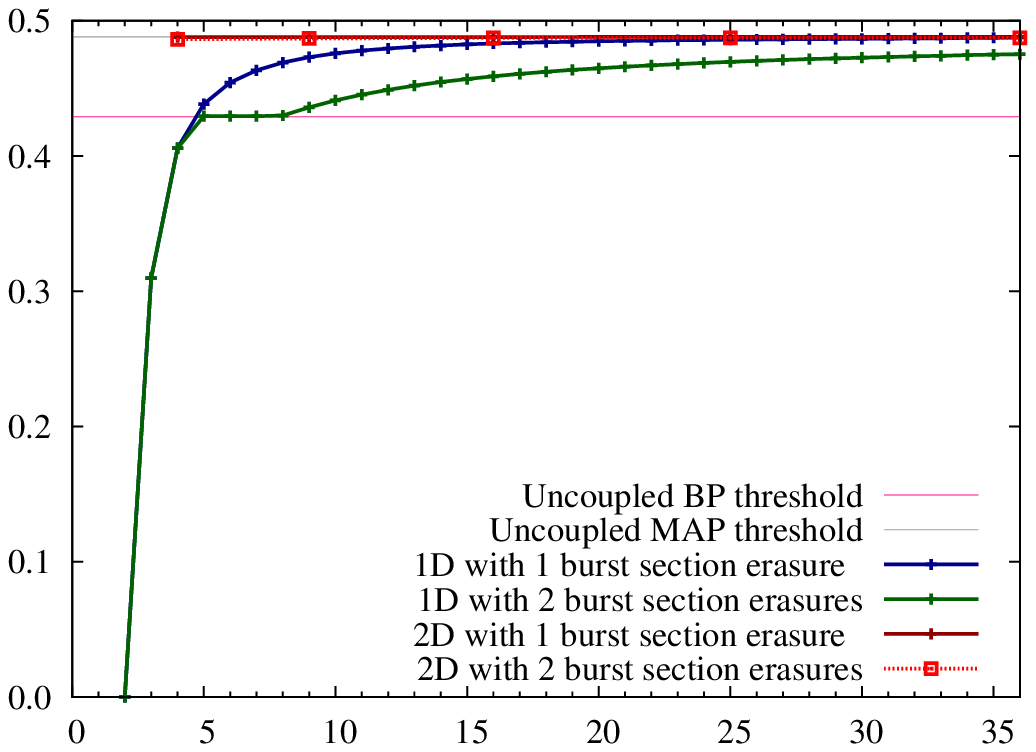}}
\put(70,0){\scriptsize{Number of coupled neighboring sections: $w^D$}}
\put(0,100){\rotatebox{90}{$\epsilon^{\ast}$}}
\end{picture} 
\caption{The BP threshold of 1D-SC and 2D-SC $(\dl=3,\dr=6,L,\omega,\ZZ)$ with $\ZZ$ as hyperplane of width $w$. The BP threshold of 1D-SC codes are badly degraded when burst section erasures exits. The 2D-SC codes are more robust than the 1D-SC codes. }
\label{012641_21Jan13}
\end{figure}

Figure \ref{012641_21Jan13} compares the BP threshold values of 1D-SC codes and 2D-SC codes with width-$w$ hyperplane shortening domain. 
The degrees $\dl$ and $\dr$  are set to 3 and 6, respectively. 
We injected one or two burst section erasures. 
The coupling number $L$ for each plotted point is chosen  sufficiently large 
so that  the BP threshold is not increased due to the rateloss, the burst error sections are not affected each other, and 
each plotted BP threshold value converges. 
Note that the BP threshold value is about 0.4882 when there is no burst section erasures. 
For small coupling window size $w\ge 4$, the BP threshold  of 1D-SC codes is 0. This is badly degraded  from 0.4882.
This degradation would not be mitigated by increasing $L$. 
When $w=2$ the BP threshold is 0, namely the burst section erasure is not recovered even if all other sections were recovered. 
This can be explained by the  theorem in the next section. 

On the other hand, 2D-SC codes are not degraded from the case of no burst section erasures even if $w$ is small. 
From this observation,  2D-SC codes are more robust to burst section erasures than 1D-SC codes. 
The x-axis indicates $w^D$. This is intended to be dealt fairly with respect to  the number of coupled neighboring sections both at 1D and 2D. 
\subsection{Bound on Performance}
In the previous section we observed that 1D-SC codes did not recover one single burst section erasures when $w=2$. 
This can be explained by the following theorem. 
\begin{theorem}
The MD $(\dl,\dr,L,\omega,\ZZ)$ code of dimension $D$ can not recover one single burst section  erasure at section $\underline{i}$ if ${\epsilon_{\underline{i}}}>
\epsilon^{\mathrm{BP}}(\dl,\dr)w^D$ if $\underline{i}\notin \ZZ$, 
where $\epsilon^{\mathrm{BP}}(\dl,\dr)$ is the BP threshold of uncoupled $(\dl,\dr)$ codes. 
\label{182307_26Jan13} 
\end{theorem}
\begin{IEEEproof}
Let us consider the best case,  namely other sections have no erasures. To be precise, $\epsilon_{\underline{j}}=0$ for $\underline{j}\neq \underline{i}$. 
The density evolution equations can be written as 
\begin{align}
p_{\underline{j}}^{(\ell)}&=0 \ (\underline{j}\neq \underline{i})\\
    p_{\underline{i}}^{(\ell)}&=\ei\Bigl(1-\sum_{\underline{j}} \omega_{\underline{j}}\bigl(1-\sum_{\underline{k}}\omega_{\underline{k}}p_{\underline{i}+\underline{j}-\underline{k}}^{(\ell-1)}\bigr)^{\dr-1}\Bigr)^{\dl-1}\\
&=\ei\Bigl(1-\bigl(1-\frac{1}{w^D}p_{\underline{i}}^{(\ell-1)}\bigr)^{\dr-1}\Bigr)^{\dl-1}.
\end{align}
Denoting $\hat{p}^{(\ell)}_{\underline{i}}:=p^{(\ell)}_{\underline{i}}/w^D$, we have
\begin{align}
& \hat{p}_{\underline{i}}^{(\ell)}=\frac{\epsilon_{\underline{i}}}{w^D}(1-(1-\hat{p}_{\underline{i}}^{(\ell-1)})^{\dr-1})^{\dl-1}.
\end{align}
This can be viewed as density evolution of uncoupled $(\dl,\dr)$ code over BEC($\epsilon_{\underline{i}}/w^D$). Hence $p_{\underline{i}}^{(\infty)}>0$ if $\frac{\epsilon_{\underline{i}}}{w^D}>\epsilon^{\mathrm{BP}}(\dl,\dr)$, which concludes the theorem. 
\end{IEEEproof}

From Theorem \ref{182307_26Jan13}, one can see that $(\dl=3,\dr=6,L,\tilde{\omega},\tilde{\ZZ})$ with $w=2$ can not recover one single burst section erasure 
since $\epsilon_{\underline{i}}=1>0.4294\times 2=0.8588=\epsilon^{\mathrm{BP}}(\dl,\dr)w^D$. 
The BP threshold gets degraded even worse when the number of burst section erasures is 2. 

\section{Rateloss Problem of Multi-Dimensional SC Codes and Its Mitigation}
As one can see in \eqref{144620_27Jan13}, the rate of SC codes is less than the uncoupled codes $1-\frac{\dl}{\dr}$. 
The 1D-SC codes have rateloss $O(1/L)$. 
The 1D-SC codes could have rateloss $O(1/L^D)$ by coupling $L^D$ sections as 1D-SC codes. 
The $D$-dimensional SC codes with hyperplane shortened domain have only $O(1/L)$ while there are $L^D$ sections. 
This is a problem. 
Is it possible to design MD-SC codes with rateloss $O(1/L^D)$ by keeping the BP threshold the same? 

Define the shortening domain as a hypercube of size $z$ as follow. 
\begin{align}
 {\ZZ}=[0,z-1]^D. 
\end{align}
We claim that the rateloss of the codes with this $\ZZ$ has rateloss $O(1/L^D)$. 
The number $C$ of check nodes that are adjacent to unshortened bit nodes is not greater than the number of all check nodes. 
\begin{align}
C\le \frac{\dl}{\dr}ML^D
\end{align}
There are $V=M(L^D-z^D)$ unshortened bit nodes. 
Thus we have the coding rate as
\begin{align}
R&=1-C/V\ge 1-\frac{\dl}{\dr}\frac{L^D}{L^D-z^D}\\
&=\Bigl(1-\frac{\dl}{\dr}\Bigr)-O\Bigl(\frac{z^D}{L^D}\Bigr). 
\end{align}
Note that we are not saying that this rate is better than the coding rate of 1D-SC codes. 
It is fair to compare the coding rate keeping the number of sections $L_C$ the same. 
From this point of view, the rateloss of both 1D-SC codes and the MD-SC codes scales with  $O(1/L_C)$, where 
$L_C=L$ for 1D-SC codes and $L_C=L^D$ for MD-SC codes of dimension $D$. 

\begin{figure}
 \begin{picture}(250,175)(0,0)
\put(10,5){\includegraphics[width=0.48\textwidth]{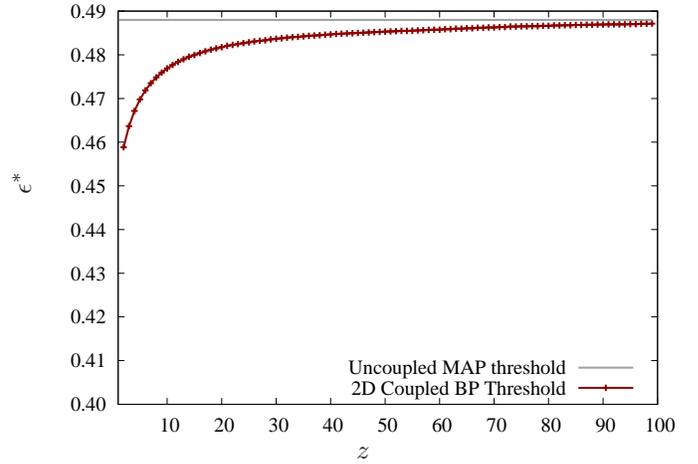}}
\put(130,0){$z$}
\put(0,100){\rotatebox{90}{$\epsilon^{\ast}$}}
\end{picture}
\caption{The BP threshold of 2D-SC $(\dl=3,\dr=6,L,w=2)$ codes with shortening domain $\ZZ$ as hypercube of size $z$. 
}
\label{152452_27Jan13}
\end{figure}

\begin{figure*}
\begin{picture}(400,270)(-10,0)
\put(  0,180){\includegraphics[width=0.245\textwidth]{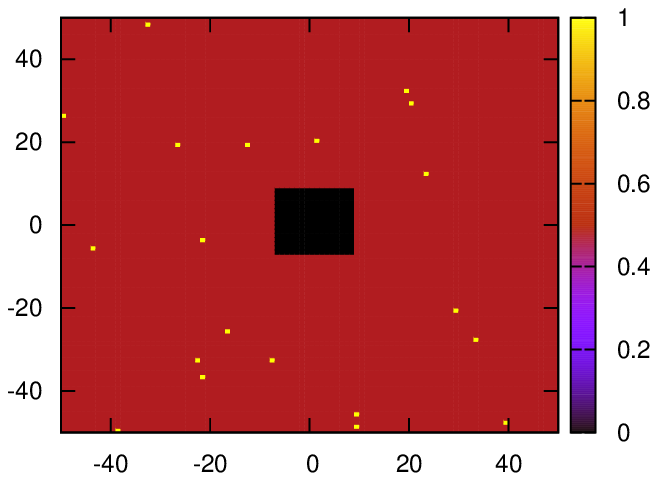}}
\put(120,180){\includegraphics[width=0.245\textwidth]{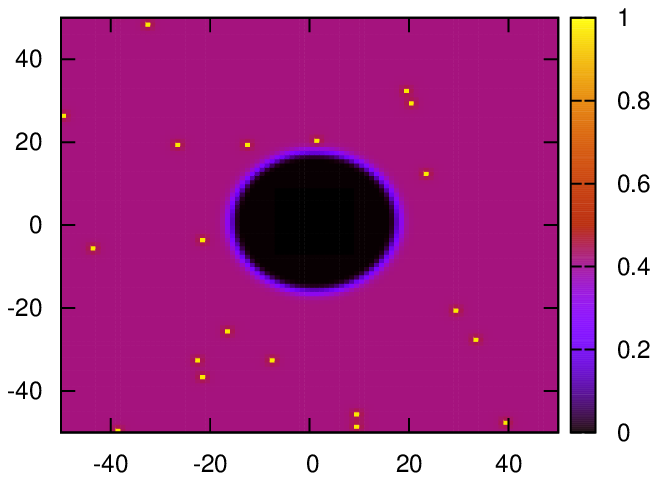}}
\put(240,180){\includegraphics[width=0.245\textwidth]{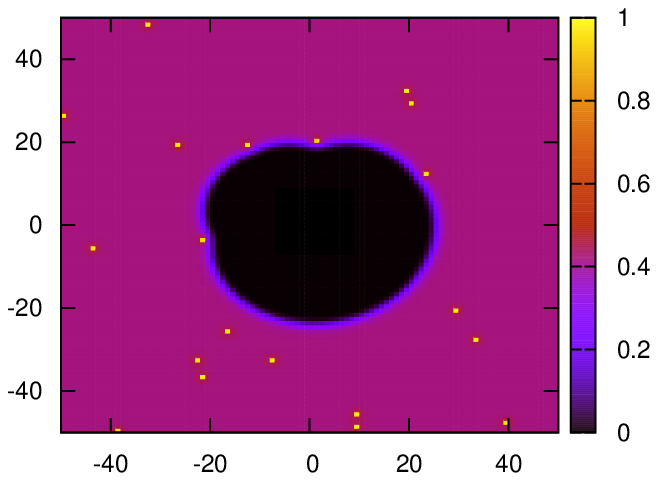}}
\put(360,180){\includegraphics[width=0.245\textwidth]{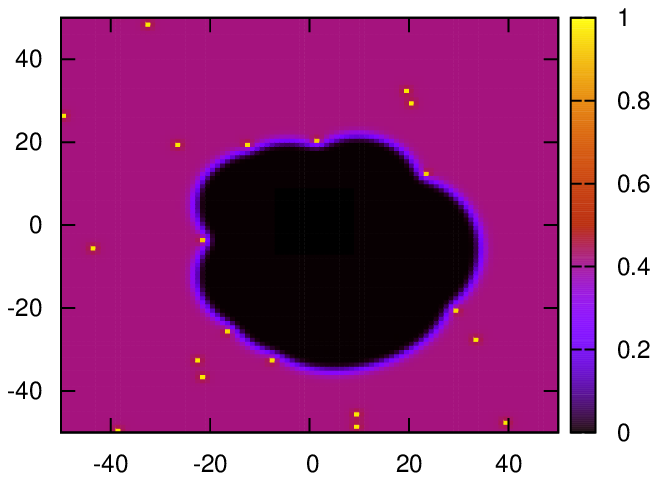}}
\put(  0, 90){\includegraphics[width=0.245\textwidth]{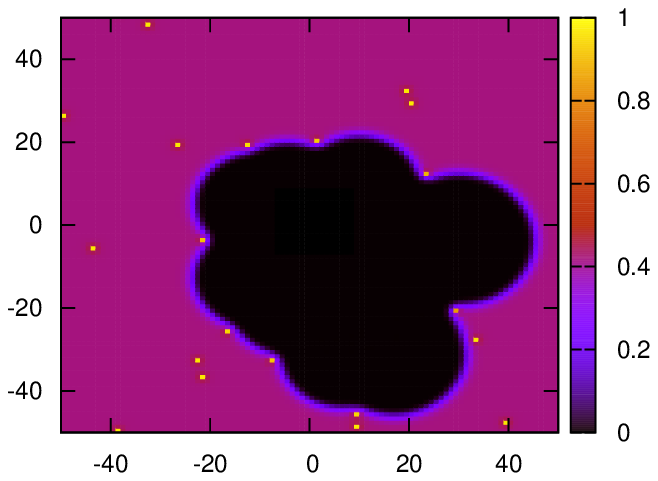}}
\put(120, 90){\includegraphics[width=0.245\textwidth]{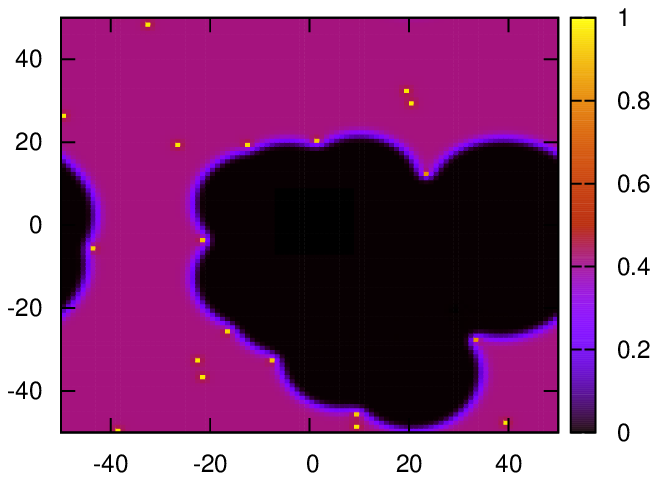}}
\put(240, 90){\includegraphics[width=0.245\textwidth]{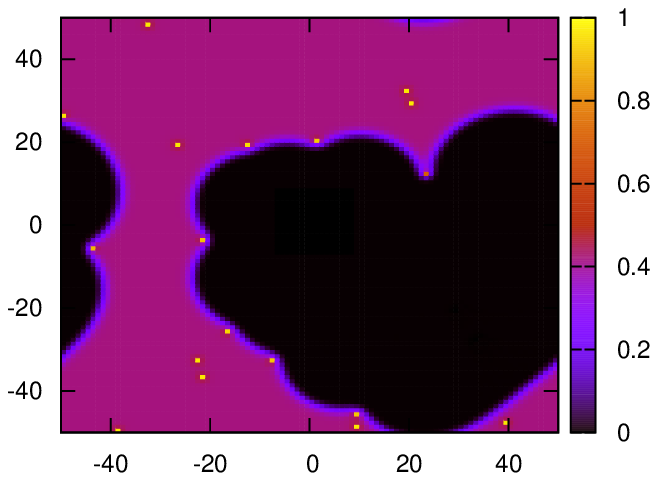}}
\put(360, 90){\includegraphics[width=0.245\textwidth]{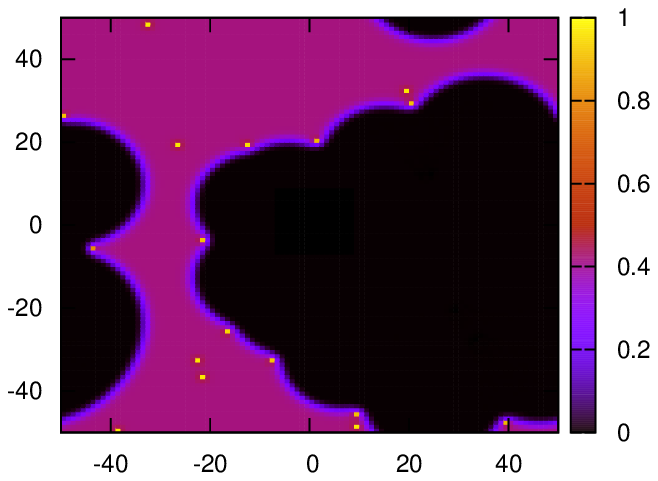}}
\put(  0,  0){\includegraphics[width=0.245\textwidth]{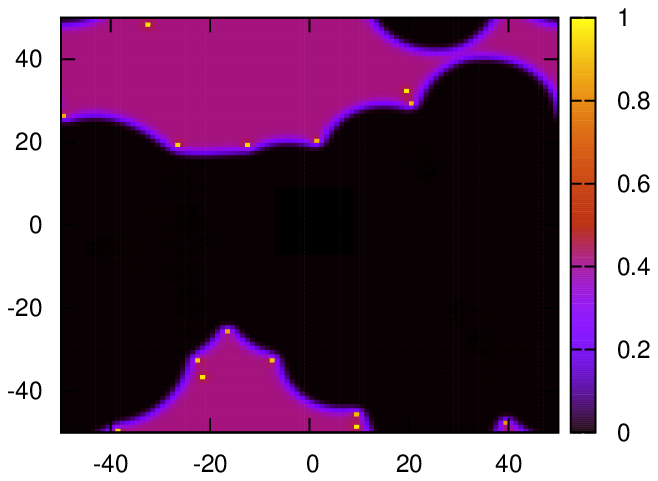}}
\put(120,  0){\includegraphics[width=0.245\textwidth]{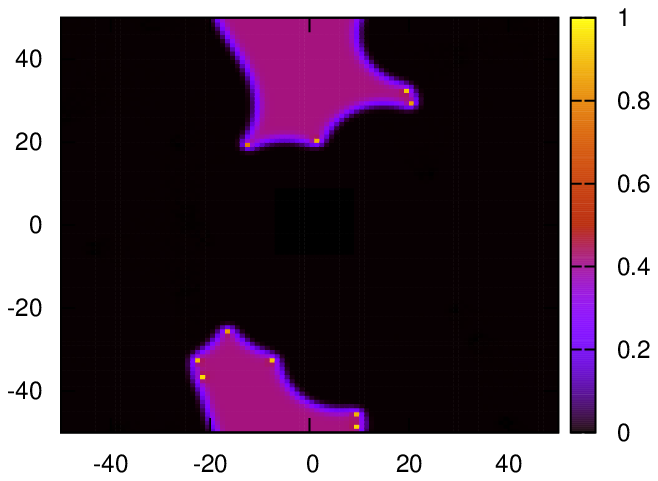}}
\put(240,  0){\includegraphics[width=0.245\textwidth]{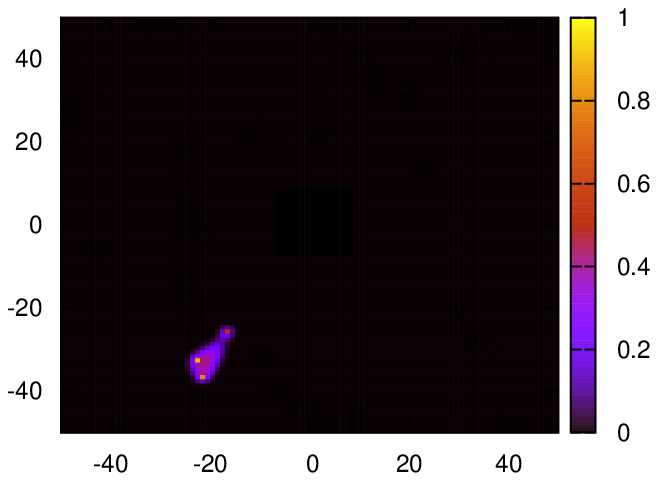}}
\put(360,  0){\includegraphics[width=0.245\textwidth]{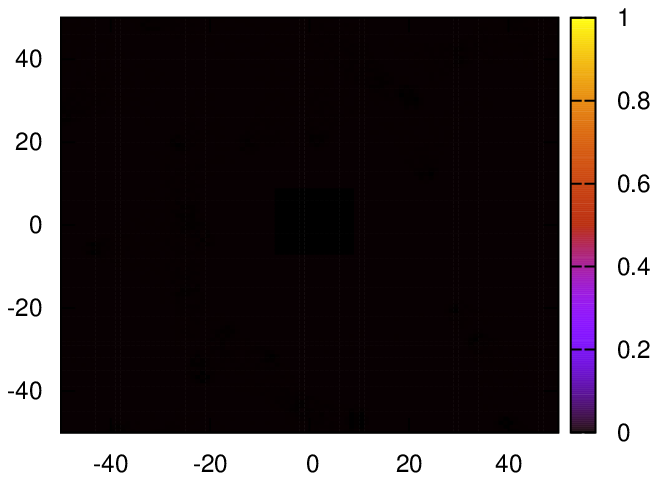}}
\end{picture} 
\caption{Transition of decoding error rate of 2D-SC $(\dl=3,\dr=6,L=101,\omega,\ZZ)$ codes with $w=2$ and  $\ZZ$ as square segment of size 15. 
 The channel is  BEC($\epsilon=0.48$) with  20 burst section erasures injected. }
\label{220826_27Jan13}
\end{figure*}

Does the BP threshold attain  the MAP threshold of the uncoupled codes?
Figure \ref{152452_27Jan13} shows the BP threshold of 2D-SC $(\dl=3,\dr=6,L,w=2)$ codes with shortening domain  $\ZZ$ as hypercube of size $z$. 
We take sufficiently large coupling number $L$ so that each plotted point converges. We observe that the BP threshold approach the MAP threshold of uncoupled codes as $z$ gets large. 
Figure \ref{220826_27Jan13} shows the transition of decoding error rate of 2D-SC $(\dl=3,\dr=6,L=101,\omega,\ZZ)$ codes with $w=2$ and  $\ZZ$ as square segment of size 15. 
 The channel is  BEC($\epsilon=0.48$) with  20 burst section erasures injected. These burst section erasures are recovered by 2D-SC codes. 
It is observed that 2D-SC codes can recover more burst section erasures as  $L$ gets large. 
\section{Conclusion}
We propose MD-SC codes. We observed that 2D-SC codes are more robust to burst section erasures than 1D-SC codes. 
\section*{Acknowledgements}
The second author would like to thank V.~Aref, N.~Macris and R.~Urbanke for helping and discussing  this work. 
The second author started this work with V.~Aref when he stayed at EPFL in 2011. 
\bibliographystyle{IEEEtran} 
\bibliography{IEEEabrv,kenta_bib}

\begin{thebibliography}{1}
\providecommand{\url}[1]{#1}
\csname url@rmstyle\endcsname
\providecommand{\newblock}{\relax}
\providecommand{\bibinfo}[2]{#2}
\providecommand\BIBentrySTDinterwordspacing{\spaceskip=0pt\relax}
\providecommand\BIBentryALTinterwordstretchfactor{4}
\providecommand\BIBentryALTinterwordspacing{\spaceskip=\fontdimen2\font plus
\BIBentryALTinterwordstretchfactor\fontdimen3\font minus
  \fontdimen4\font\relax}
\providecommand\BIBforeignlanguage[2]{{%
\expandafter\ifx\csname l@#1\endcsname\relax
\typeout{** WARNING: IEEEtran.bst: No hyphenation pattern has been}%
\typeout{** loaded for the language `#1'. Using the pattern for}%
\typeout{** the default language instead.}%
\else
\language=\csname l@#1\endcsname
\fi
#2}}

\bibitem{zigangirov99}
A.~J. Felstr{\"o}m and K.~S. Zigangirov, ``Time-varying periodic convolutional
  codes with low-density parity-check matrix,'' \emph{{IEEE} Trans. Inf.
  Theory}, vol.~45, no.~6, pp. 2181--2191, June 1999.

\bibitem{5571910}
M.~Lentmaier, A.~Sridharan, D.~Costello, and K.~Zigangirov, ``Iterative
  decoding threshold analysis for {LDPC} convolutional codes,'' \emph{{IEEE}
  Trans. Inf. Theory}, vol.~56, no.~10, pp. 5274--5289, Oct. 2010.

\bibitem{5695130}
S.~Kudekar, T.~Richardson, and R.~Urbanke, ``Threshold saturation via spatial
  coupling: Why convolutional {LDPC} ensembles perform so well over the
  {BEC},'' \emph{{IEEE} Trans. Inf. Theory}, vol.~57, no.~2, pp. 803--834, Feb.
  2011.

\bibitem{2012arXiv1201.2999K}
S.~{Kudekar}, T.~{Richardson}, and R.~{Urbanke}, ``{Spatially Coupled Ensembles
  Universally Achieve Capacity under Belief Propagation},'' \emph{ArXiv
  e-prints}, Jan. 2012.

\end{thebibliography}
\end{document}